\documentclass[aps,twocolumn,groupedaddress,superscriptaddress,amsmath,amssymb,prl]{revtex4-2}

\usepackage{comment}
\usepackage{amsfonts}
\usepackage{amsmath}
\usepackage{float}
\usepackage{physics}
\usepackage{mathtools}
\usepackage{bbold}
\usepackage{amsthm}
\usepackage[margin=25mm]{geometry}
\usepackage[table,xcdraw]{xcolor}
\usepackage{amsmath}
\usepackage{braket}
\usepackage{hyperref}
\usepackage{bm}
\usepackage{tikz}
\usepackage{pgfplots}
\usepackage{tikzscale}
\usepackage{ mathrsfs }
\newcommand{\bqa}{\begin{eqnarray}}
\newcommand{\eqa}{\end{eqnarray}}
\usepackage[export]{adjustbox}
\usepackage[normalem]{ulem}
\usepackage{placeins}

\usepackage{caption}
\usepackage{subcaption}
\newcommand{\be}{\begin{equation}}
\newcommand{\ee}{\end{equation} }
\newcommand{\new}[1]{{\color[rgb]{0.0,0,0.6}{#1}}}

\newcommand{\beginsupplement}{%
        \setcounter{table}{0}
        \renewcommand{\thetable}{S\arabic{table}}%
        \setcounter{figure}{0}
        \renewcommand{\thefigure}{S\arabic{figure}}%
        \setcounter{equation}{0}
        \renewcommand{\theequation}{S\arabic{equation}}%
     }

\newcommand{\Imperial}{Blackett Laboratory, Imperial College London, SW7 2AZ, United Kingdom}
\newcommand{\HZDR}{Helmholtz-Zentrum Dresden-Rossendorf, Bautzner Landstraße 400, 01328 Dresden, Germany}
\newcommand{\sussex}{Sussex Centre for Quantum Technologies, University of Sussex, Brighton, BN1 9RH, United Kingdom}

\newcommand{\uq}{Universal Quantum Ltd, Haywards Heath, RH16 1XQ, United Kingdom}

\begin{document}

\title{Amplitude-noise-resilient entangling gates for trapped ions}

%Amplitude-noise resilience of trapped-ion entangling gates}
%Robust entangling gate for two ions in an anharmonic trap}

\author{Nguyen H. Le}\affiliation{\Imperial}
\author{Modesto Orozco-Ruiz}\affiliation{\Imperial}
\author{Sahra A. Kulmiya}\affiliation{\sussex}\affiliation{\uq}
\author{James G. Urquhart}\affiliation{\sussex}
\author{Samuel J. Hile}\affiliation{\sussex}
\author{Winfried K. Hensinger}\affiliation{\sussex}\affiliation{\uq}
\author{Florian Mintert}\affiliation{\Imperial}\affiliation{\HZDR}

%\author{abc}
%\affiliation{def}

\begin{abstract}
Noise resilience of quantum information processing is a crucial precondition to reach the fault-tolerance threshold.
While resilience to many types of noise can be achieved through suitable control schemes, resilience to amplitude noise seems to be elusive within the common harmonic approximation for the bus mode of trapped ions. We show that weak an-harmonicities admit control schemes that achieve amplitude noise-resilience consistent with state-of-the-art experimental requirements,
and that the required an-harmonicities can be achieved with current standards of micro-structured traps or even the intrinsically an-harmonic Coulomb interaction. This approach applies broadly to any platform that employs a bosonic bus as a qubit coupler.
% We show that a small amount of anharmonicity in the trapping potential can be utilized for implementing a robust Molmer-Sorensen gate which exhibits resilience against amplitude variations. The anharmonicity changes the geometric nature of the gate dynamics from two-dimensional translations in phase space to multi-dimensional rotations of a qudit. We investigate the specific case of two trapped Yb ions and derive optimal pulses for achieving 99.99\% fidelity even in the presence of amplitude uncertainties of up to 10\%. The required anharmonicity can be induced by either higher orders in the trapping potential or the nonlinear Coulomb repulsion between the ions in the breathing mode. 
\end{abstract} 
\maketitle

Bosonic couplers are ubiquitous in emerging quantum computing platforms. By utilizing quantized excitation of a bosonic mode such as phonon in a collective motional mode \cite{cirac_quantum_1995,MS1999}, photon in a cavity field \cite{welte_photon_2018,Majer2007,dicarlo_demonstration_2009,Allen2017,srinivasa_cavity-mediated_2024,dijkema_cavity-mediated_2024}, or collective excitation in superconducting circuit \cite{yan_tunable_2018,sung_realization_2021}, it is possible to engineer a quantum bus that mediates interaction between qubits. %Entanglement can be generated by direct or effective driving of the bus \cite{yan_tunable_2018, Allen2017, welte_photon_2018,MS1999}. 
While proof of principle demonstrations of entangling gates exist for a variety of platforms from trapped ions \cite{Leibfried2003,Webb2018}, neutral atoms \cite{welte_photon_2018}, cavity-coupled spins \cite{dijkema_cavity-mediated_2024,Harvey2022,Borjans2020}, to superconducting qubits \cite{sung_realization_2021,dicarlo_demonstration_2009,google_ai_2020,Majer2007},
any practical device will require a certain level of noise-resilience so that the benefits of using a quantum instead of a classical computer are not lost to the effort required for highly accurate and frequent system calibration \cite{wittler_integrated_2021,tornow_minimum_2022,gerster_2022,majumder_2020}.

For trapped ions, existing demonstrations of resilience of %trapped-ion 
quantum gates against fluctuations of a variety of quantities \cite{Webb2018, milne2020phase,Zarantonello2019,Hayes2012, shapira2018robust, bermudez2012robust} put this architecture much closer to practicality than many competing platforms. By design, most of the currently employed quantum gates are resilient to fluctuations in the initial state of the ions' motion~\cite{MS1999, sorensen_entanglement_2000,Leibfried2003}.
Resilience against motional heating and fluctuations in the ions' confining potential or carrier frequency of driving fields used to realize gates can be achieved in term of suitably tailored temporal shapes of the driving fields \cite{Lishman2020,Orozco2024,Haddadfarshi2016, bermudez2012robust, milne2020phase}.
A crucial system parameter that has proven tricky to achieve noise resilience against is the amplitude of driving fields;
but typical fluctuations in Rabi-frequency are in the range of a few percents.  

The linear spatial dynamics of the ions, is conflicting with resilience against amplitude fluctuations of driving fields.
The required nonlinearity can be obtained from the intrinsically nonlinear light-matter interaction~\cite{Shapira2023} beyond the Lamb-Dicke approximation. As we will show here, it is possible to achieve resilience against amplitude fluctuations without the strong driving required for sizeable nonlinearity in the light-matter interaction using anharmonicities in trapping potential or even the fundamentally anharmonic Coulomb interaction.
Even though such anharmonicities impair resilience to thermal excitations, they do so only to an extent that can be reclaimed with the choice of temporal profile of the driving fields.
With the explicit design of gate electrodes, we underpin the experimental feasibility of the required anharmonic potential.
Although we present the gate scheme specifically for trapped ions, it can be extended to any architecture that uses a weakly anharmonic (or weakly nonlinear) bosonic mode as a bus. In such setups, this scheme mitigates detrimental effects of amplitude fluctuations and unwanted excitations in the coupler.

The Hamiltonian of a pair of trapped ions with off-resonant driving on a red and a blue sideband is given by \cite{MS1999}
\begin{equation}
\label{eq:Hc}
H(t)=H_0+\Omega_R\left(f(t)a^{\dagger}+af^\ast (t)\right)S_y\ ,
\end{equation}
where 
$S_y=\sigma_y^1+\sigma_y^2$ is the $y$--component of the total spin operator,
$a$ and $a^\dagger$ are the annihilation and creation operator of the bus mode;
$\Omega_R$ is the Rabi frequency for the utilized side-band transitions, and the function $f(t)$ includes the time-dependence of the carrier frequencies of the driving fields and any time-dependence resultant from pulse shaping.
$H_0$ is the non-interacting Hamiltonian of the internal, qubit degrees of freedom of the ions and the bus mode.
 
In the case of a perfectly harmonic bus mode, the impact of the non-interacting part $H_0$ of the system Hamiltonian reduces to an oscillatory time-dependence of the annihilation and creation operators $a$ and $a^\dagger$.
The system Hamiltonian in the interaction picture thus reads
$\tilde H(t)=\Omega_R(\tilde f(t)a^{\dagger}+a\tilde f^\ast (t))S_y$ with a drive%driving function 
$\tilde f(t)$ dressed with the time-dependence of the non-interacting dynamics.

The gate dynamics can be represented by translation in phase space along a closed loop with length proportional to $\Omega_R$ \cite{MS1999,sorensen_entanglement_2000}.
The Rabi-angle $\Phi_R$ of the effective $S_y^2$--interaction in the dynamics induced by $\tilde H(t)$ is given by the area enclosed by the loop, thus exhibiting the quadratic dependence
$\Phi_R=\Omega_R^2  \Im \int_{0}^T d \tau  \tilde f(\tau) \int_{0}^{\tau}d \tau^\prime \tilde f^\ast(\tau^\prime)$~\cite{Lishman2020}.
The dependence of $\Phi_R$ on the driving pattern  factorizes into an amplitude term $\Omega_R^2$ and a factor with the detailed time-dependence of the driving.
There is thus no possibility of choosing driving patterns $f(t)$ that could modify the quadratic dependence on the Rabi-frequency and any fluctuation of $\Omega_R$ will unavoidably result in the corresponding fluctuation of the Rabi-angle $\Phi_R$. 

In the case of an anharmonic bus mode, however, the interplay between the interaction and the non-interacting dynamics can break this factorization, and it is possible to achieve resilience against fluctuations in the Rabi frequency $\Omega_R$ in terms of suitably tailored driving patterns $f(t)$.

The ideal entangling gate for the qubit degrees of freedom that can be realised with the Hamiltonian $H(t)$ (Eq.~\eqref{eq:Hc}) is given by
\be\label{eq:UT}
U_T=\exp\left(i\frac{\pi}{8}S_y^2\right)\ .
\ee
Since any level of anharmonicity will restrict a gate functionality to a limited range of initial motional states, it is essential to define a gate fidelity for the joint dynamics $V$ of qubits and bus mode that takes into account this range.
For any projector $P$ onto a subspace of the full Hilbert space of the bus mode, one can define
\be\label{eq:F}
F(V,U_T,P)=\left\vert\mbox{tr}((U_T^{\dagger}\otimes P)V)/(4\tr P)\right \vert^2,
\ee
%\flo{should we use $V$ for unitaries of qubits and bus?}
as the fidelity of a unitary $V$ for the full system of qubits and bus mode with respect to the desired gate $U_T$ of the qubits and the desired trivial dynamics of the bus mode within the subspace given by $P$.
Resilience against amplitude fluctuations is characterized in terms of an averaged infidelity
\be
I=1-\langle F(V(\Omega_R),U_T,P)\rangle_{\Omega_R}\ ,
\label{eq:ensembleF}
\ee
where $V(\Omega_R)$ is the system dynamics obtained with a given Rabi frequency $\Omega_R$ taking value in the error range $\Omega_C - \delta \Omega\leq \Omega_R\leq \Omega_C+\delta \Omega$ where $\Omega_C$ is the central value and $\delta \Omega$ the error magnitude.

Although the anharmonicity can be induced by any higher order terms in the potential, for clarity the remaining discussion focuses on the quartic potential $\frac{1}{2}m\omega^2\left(z^2+z^4/\xi^2\right)$ for the bus mode, where $\xi$ is the length scale on which the potential becomes anharmonic.
The perturbative correction to the center of mass (COM) mode's eigen-frequencies resultant from the anharmonicity is given by $\chi n(n-1)$ with the phonon number $n$ and the scalar prefactor
\be
\chi=\frac{3\hbar}{4 m \xi^2}
\ee
referred to as the {\em anharmonicity} in the following.

\begin{figure*}[t]
\includegraphics[width=2\columnwidth]{fig_infid_5pc.pdf}
\caption{ (a) Dependence of infidelity on Rabi frequency's variation for the Molmer-Sorensen (MS) gate in a harmonic trap (solid red), and for the anharmonic gate with 1 phonon excitation (with $\chi=2.5\Omega_G, \Omega_C  =2.3\Omega_G$, dotted blue), and anharmonic gate with 10 phonon excitations (with $\chi=6 \Omega_G, \Omega_R =2.6 \Omega_G$, dashed orange); here $\Omega_G=2\pi/T$ is the gate frequency.
(b) Infidelity averaged over a 5\% error range of the Rabi frequency for 1 phonon excitation and (c) up to 10 phonon excitations, as a function of the anharmonicity and mean Rabi frequency. \label{fig:fidmap}}
\end{figure*}

With the anharmonic potential terms included in the Hamiltonian of Eq.~\eqref{eq:Hc}, the driving function $f(t)$ is optimized with common pulse-shaping algorithms~\cite{khaneja2005optimal, borzi2008formulation, de_fouquieres_second_2011}.  The optimization is performed over an ensemble of Hamiltonians that share the same general form but differ in their Rabi frequencies, $\Omega_R$. The goal is to design a single driving function $f(t)$ that works effectively for all members of the ensemble, ensuring consistent control despite variations in $\Omega_R$. In the following, we will pursue a numerically exact approach (given $H(t)$ in Eq.~\eqref{eq:Hc}), applicable to any anharmonicity, and an approximate analytic approach that is applicable to the regime of strong anharmonicities.
The latter approach provides an intuitive understanding of the functionality of the control scheme, and
using its driving functions as initial condition for the iterative refinement of the former approach helps to avoid sub-optimal extrema.

Fig.~\ref{fig:fidmap}a depicts the infidelity $1-F$ (Eq.~\eqref{eq:F}) for several gates as function of the Rabi frequeny $\Omega_R$.
The solid red curve corresponds to a perfectly harmonic system in which no resilience can be achieved,
and the infidelity grows quickly with increasing deviation $|\Omega_R-\Omega_C|$ of the Rabi frequency from its ideal value.
The other two curves corresponds to an anharmonic system, showing the fidelities obtained with driving patterns optimized for an equally spaced grid of Rabi frequencies in the interval $[9.5/10\ \Omega_C,10.5/10\ \Omega_C]$ centered around the central Rabi frequency $\Omega_C$. The pulses are optimised under the constraint $|f(t)|\le 1$. The dotted blue refers to gate fidelities with up to 1 phonon in the initial states, i.e., $P=\dyad{0}+\dyad{1}$ in Eq.~\eqref{eq:F}, for an anharmonicity $\chi=2.5\Omega_G$ where $\Omega_G=2\pi/T$ is  the gate frequency with $T$ the gate duration.
The dashed orange refers to the case with up to 10 phonons, $P=\sum_{n=0}^{10}\dyad{n}$, and anharmonicity $\chi=6\Omega_G$.
The value of the gate frequency is chosen to be $\Omega_G=\Omega_C/2.3$ for the former gate;
given the broad range of initial states, the latter gate requires a slightly lower gate frequeny $\Omega_G=\Omega_C/2.6$ in order to achieve the resilience shown in Fig.~\ref{fig:fidmap}a.

A more quantitative picture of the noise-resilience is provided by the average infidelity $I$ (Eq.~\eqref{eq:ensembleF}). %assessed with an ensemble of Rabi-frequencies not used for the optimization.
Figures ~\ref{fig:fidmap} b and c depict $I$ 
%for ensembles drawn from the same uniform distribution as above 
as function of the central Rabi frequency $\Omega_C$ and the anharmonicity $\chi$ for the cases of up to 1 phonon and up to 10 phonons in the initial states, respectively.

Both insets show sub-optimal infidelities for vanishing anharmonicity, as expected.
With increasing anharmonicity, however, the infidelities decrease
and values below $10^{-4}$ are achieved in the strong anharmonicity limit.
This decrease is faster in inset~\new{(b)} than in inset~\new{(c)}, highlighting that the required anharmonicity increases with the dimension of the subspace to which its initial motional state is confined.
While a Rabi frequency of value $\Omega_R=\Omega_G$
is enough to realize a fully entangling gate in the absence of amplitude fluctuations, Fig.~\ref{fig:fidmap} b and c also show that -- depending on the desired infidelity -- slightly larger Rabi-frequencies can be required in order to realize resilient gates.
%We found that 
In addition, the required anharmonicity for achieving an average fidelity of 99.9\% rises moderately with error magnitude for error values in the range 1\%-10\% (Supp. Mat. Sec. 4). While robustness against motional excitation is demonstrated for up to 10 phonons in Fig.~\ref{fig:fidmap}c, to highlight the potential of the method, typical motional excitation levels in many experiments are significantly lower, with $\langle n \rangle \ll 1$~\cite{benhelm_towards_2008,gaebler_high_2016,schafer_fast_2018}. In this regime of low motional excitation, higher gate fidelities can be achieved with weaker anharmonicity, as illustrated in Fig.~\ref{fig:fidmap}b.

In order to understand the physical origin of the robustness against fluctuations in the Rabi frequency $\Omega_R$, it is instructive to pursue an approximate treatment  that is valid for large anharmonicity,
a regime, in which a transition between any pair Fock states can be driven on resonance without sizeable off-resonant transitions between other pairs.
%$\ket{n}$ 
%\nguyen{$\ket{n-1}$} and %$\ket{n+1}$
%\nguyen{$\ket{n}$} can be driven without driving any other transition.
The realization of a gate that works with an initial motional state in the subspace spanned by the lowest $N$ Fock states requires a driving profile with components $g_n(t)$ close-to-resonant with the transition between one pair of Fock states each (Eq.~\eqref{eq:polycontrol} in Supp. Mat.).

With such a driving profile, the system Hamiltonian in the interaction picture is approximated (in rotating wave approximation) as 
\be\label{eq:Hrwa}
H_I=\Omega_R\!\sum_{n=1}^{N}\sqrt{n}\Bigl(g_n(t)\sigma_n^\dagger+\sigma_{n}g_n^\ast(t)\Bigr)S_y,
\ee
with $\sigma_{n}=\dyad{n-1}{n}$.
The operators
\begin{subequations}
\be
X_{n}=\frac{1}{2}(\sigma_n+\sigma_n^\dagger)S_y\ ,\mbox{ and }
Y_{n}=\frac{i}{2}(\sigma_n-\sigma_n^\dagger)S_y
\ee
satisfy the commutation relation 
$[X_{n},Y_{n}]=2iZ_n$  with
$
Z_n=[\sigma_n^\dagger,\sigma_n]S_y^2/4\,
$
\end{subequations}
and cyclic permutations, %as $S_y^3=4S_y$,
{\it i.e.} the same commutation relations as the Pauli operators.
The dynamics with driving of a single transition is thus equivalent to that of a single qubit.

If the dynamics $V$ of qubit degrees of freedom and bus mode satisfy the relation
\be
V(\bm{1}\otimes P)=\exp\left(-i\frac{\pi}{2}\sum_{n=1}^{N} n Z_n\right)(\bm{1}\otimes P)\ ,
\ee
then the desired gate for initial motional states in the subspace $P=\sum_{n=0}^{N-1}\ket{n}\bra{n}$ is realised,
and thus the infidelity $I$ (Eq.~\eqref{eq:F}) is minimized.
Such dynamics can be realised in terms of a sequence of steps in which a single transition is driven with a driving function $g_n$ that is optimized for the target dynamics {$\exp\left(-in\pi/2\, Z_n\right)$, or any other target within $2\pi$-periodicity.
In fact, since the dynamics resultant from driving profile $g_n(t)$ commutes with the dynamics resultant from driving profile $g_m(t)$ for $|n-m|>1$, such a driving scheme can be comprised of two steps only with $g_n=0$ for all even $n$ in one step, and $g_n=0$ for all odd $n$ in the other step.

\begin{figure}[t]
\begin{subfigure}[t]{0.22\textwidth}
\includegraphics[width=\columnwidth,left]{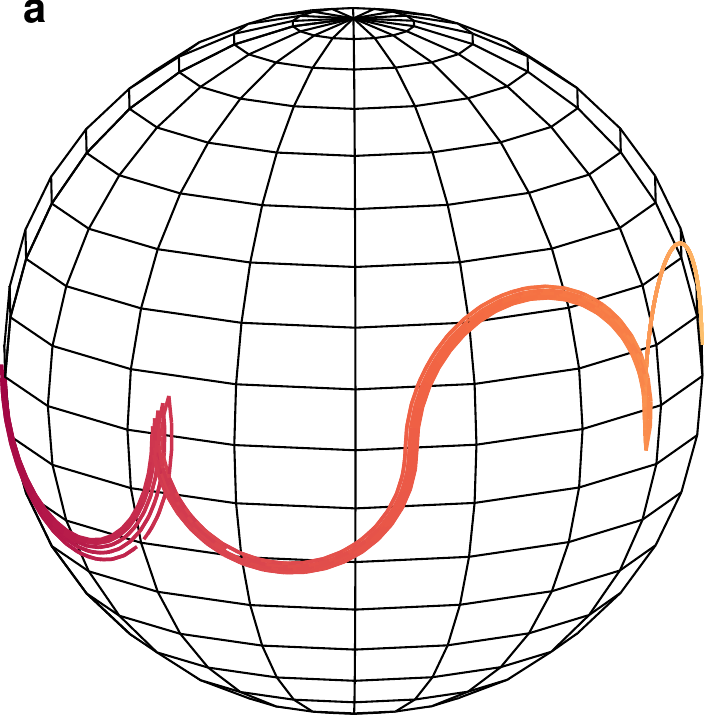}
\end{subfigure}
\hfill
\begin{subfigure}[t]{0.22\textwidth}
\centering\includegraphics[width=\columnwidth,right]{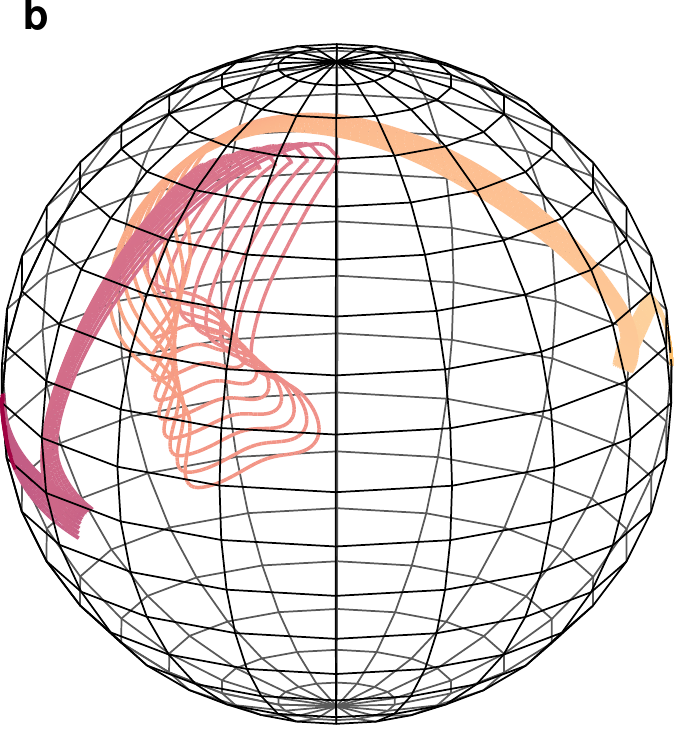}
\end{subfigure}
\caption{(a) The trajectories of the Bloch vector in the strong anharmonicity limit with the analytic pulse sequence of Eq.~\eqref{eq:piecewiseconstant}, for  Rabi frequency varying in a 10\% error range.
The color gradient depicts temporal evolution from bright orange for the initial vector $[1,0,0]$ to dark red for the final vector $[-1,0,0]$.
(b) Similar trajectories for a finite anharmonicity, $\chi =2 \Omega_G$, and a 10\% error range in the Rabi frequency. The trajectories shown are projections in the subspace of the lowest two levels, thus residing inside the Bloch sphere. The opacity represents the length of the projected vector.}
\label{fig:bloch} 
\end{figure}

For a two-level transition the unitary dynamics can be made robust against pulse amplitude error using composite pulses \cite{levitt_composite_1986,cummins_tackling_2003,genov_correction_2014}. A possible choice for each of the driving functions $g_n(t)$ that achieves the desired resilience to amplitude noise is given by the simple piecewise constant driving function with four segments $\frak{g}_j$, for $(j-1)T/4\leq t<jT/4$~\cite{cummins_tackling_2003}, with
\bqa
\frak{g}_1=\frak{g}_4^\ast &=&\frac{2\pi i}{\sqrt{n}\,\Omega_C T}\exp(-i\frac{3\phi}{4})\ ,\nonumber \\
\frak{g}_2=\frak{g}_3^\ast &=&\frac{2\pi i}{\sqrt{n}\,\Omega_C T}\exp(-i\frac{\phi}{4})\ .
\label{eq:piecewiseconstant}
\eqa

This driving pattern for $g_n(t)$ results in the gate $\exp(i\phi Z_{n})$
at the final time $T$ given a central Rabi-frequency with the value $\Omega_C$.
Fluctuations in the Rabi-frequency contribute only quadratically to the gate angle and deviations from the type of gate ({\it i.e.} induced by $Z_{n}$) are of third order in Rabi-frequency fluctuations.
That is, the gate is resilient to amplitude fluctuations up to second order, resulting in a robustness up to fourth order in the gate fidelity.

Fig.~\ref{fig:bloch}a depicts trajectories on the Bloch sphere (defined in terms of %$X_{n}$, $Y_{n}$ and $Z_{n}$  
$X_{1}$, $Y_{1}$ and $Z_{1}$  for the dynamics induced by the analytic pulse sequence in Eq.~\eqref{eq:piecewiseconstant} with $\phi=\pi/2$.
Trajectories for relative Rabi frequencies $\Omega_R/\Omega_C$ from $0.9$ to $1.1$ in steps of $0.025$ are shown. The color gradient indicates the temporal evolution with bright orange for $t=0$ to dark red for $t=T$.
The trajectories diverge at first due to variation in the Rabi frequency, but converge toward the end, demonstrating the robustness of the gate. Outside the regime of strong anharmonicity and weak driving, the separation into dynamics in distinct two-dimensional subspaces breaks down. Yet, a three-dimensional projection of the $N^2-1$ generalized Bloch vector (onto $X_{1}$, $Y_{1}$ and $Z_{1}$), shown in Fig.~\ref{fig:bloch}b for $\chi=2\Omega_G$, can still reveal the refocusing of the trajectories towards the gate time.

The anharmonicity necessary for the present scheme can be induced by a quartic potential in a trap geometry where the DC control electrodes are placed directly underneath the ions~\cite{Nizamani2012}. It scales as the inverse of the square of the separation between the electrodes and thus can be enhanced by reducing the size of the device.  We find that  it is more advantageous to use the stretch mode, rather than the COM mode, as the intrinsic anharmonic Coulomb interaction can produce a substantial contribution to the anharmonicity. As discussed in more detail in Supp. Mat. Sec. 1,
anharmonicities around $2\pi\times 100$Hz are readily achievable via the intrinsic Coulomb interaction, and values exceeding $2\pi \times 1$kHz can be obtained with the addition of a  quartic component with $\xi \sim 1 \mu$m in the trap potential. While anharmonicity in the trap potential may lead to coupling of modes \cite{home2011} which causes loss of fidelity \cite{Zhu2006,Zhu2006PRL,Wu2018}, the anharmonicity required in our gate is much smaller than typical trap frequency, in the range $ \omega/2\pi \sim 0.1 -  1$MHz, and thus
the coupling between different motional modes 
is well negligible, so that the reduction of the motional dynamics to only the bus mode is well justified. Changes to the coherence properties of the motional bus due to such a small anharmonicity are also insignificant.

The required anharmonicity for achieving a 99.9\% average fidelity in the present scheme is on the order of $\Omega_G\equiv 2\pi/T$ (see Supp. Mat. Sec. 4). A gate time on the order of 1ms is thus achievable for anharmonicity on the order of  $\chi/2\pi \simeq 1$kHz. Although the corresponding gate time of $1$ms is longer than the typical duration  for the MS gate ($\sim 100 \mu$s) \cite{ballance_high_2016}, it is much shorter than the typical coherence time on the order of minutes \cite{harty_high_2014,haffner_robust_2005}, and hence the effect of decoherence is negligible.

An appealing feature of anharmonicity-based approaches is that noise resilience can be achieved with a Rabi frequency comparable to that of an MS gate (as seen in Fig.~1b). In contrast, approaches based on light-matter nonlinear interaction, where the nonlinearity is second-order in the Lamb-Dicke parameter \cite{Shapira2023}, require orders-of-magnitude increases in the Rabi frequency to compensate for the small nonlinearity in systems with small Lamb-Dicke parameters.

The present scheme can be applied for two selectively driven ions in a multi-ion chain. As with schemes based on higher-order nonlinearity \cite{Shapira2023}, the anharmonicity decreases with the number of ion as $1/n$ (Supp. Mat. Sec. 4). However,  most scalable trapped-ion  architectures utilize one-qubit and two-qubit gates in separate entanglement zones, each with only a few ions, and interleave the gates with ion transport between zones \cite{hensinger_t-junction_2006,kielpinski_architecture_2002,pino_demonstration_2021}, thus,  these noise-resilience schemes are still applicable.

While they are discussed here for the specific platform of trapped ions, both the problem of amplitude fluctuations and the foundations of the presently proposed solution are prevalent in many quantum technological platforms:
interactions between superconducting qubits for example can be mediated via weakly anharmonic qubit couplers~\cite{yan_tunable_2018,sung_realization_2021,google_ai_2020} and long-range interactions are frequently realized via coupling to a shared cavity mode \cite{Majer2007,Harvey2022,Borjans2020}. With intrinsic or engineered anharmonicities, all such systems can benefit from the noise-resilience that can be achieved with control techniques following the principles exemplified here with the specific example of trapped ions.
The present techniques thus do not only help to bring trapped ion technology closer to the error-correction threshold, but they can find application in a broad platform of emerging technologies.

\section{Acknowledgements}
This work was supported by the U.K. Engineering and Physical Sciences Research Council via the EPSRC Hub in Quantum Computing and Simulation (EP/T001062/1), the UK Innovate UK (project number 10004857), 
the US Army Research Office (W911NF21-1-0240), the US Office of Naval Research under Agreement No. N62909-19-1-2116. S.A.K. acknowledges support from an EPSRC Centre for Doctoral Training (EP/S023607/1).
\bibliography{robust.bib}
\newpage
\onecolumngrid
\beginsupplement
\section*{Supplementary Materials}
\subsection{1. Anharmonicity estimation}
Here we estimate the anharmonicty for the COM mode and the stretch mode in the quartic potential
\be
W(z)=\frac{m\omega^2}{2}\left(z^2+\frac{z^4}{\xi^2}\right).
\ee
For the COM mode, the effective potential is the same as that of a single trapped ion. The level shift is estimated by perturbation theory $m\omega^2\bra{n}z^4\ket{n}/2\xi^2$, giving $(n^2+n)\hbar \chi$ where $\chi=3\hbar/(4m\xi^2)$. We write this as $\hbar\Delta_n+2n\hbar\chi$ where $\Delta_n=(n^2-n)\chi$ and group the second term into the harmonic part of the energy spectrum, redefining $\omega$ as the transition between the shifted ground and first excited levels.

In addition to the energy shift, the quartic potential also gives rise to cross terms of the form $\dyad{n}{n+2}+h.c.$ and $\dyad{n}{n+4}+h.c.$ in the lab frame. The dynamics of the entangling gate, however, is considered in the rotating frame of the harmonic part of the motional Hamiltonian, $\omega (n+1/2)$, and the cross terms become rapidly oscillating with frequencies $2\omega$ and $4\omega$, which are much larger than the relevant energy scale of the gate dynamics, i.e., the Rabi frequency and the anharmonic energy shift. Therefore, these cross terms can be neglected in the rotating wave approximation. The contribution of these fast oscillating terms is on the order of $\chi/\omega$ in the unitary evolution operator, and $(\chi/\omega)^2$ in the infidelity, which is in the range of $10^{-6}-10^{-4}$ for the parameter values considered in the paper, $\chi/2\pi \sim 1$kHz and $\omega/2\pi \sim 0.1-1$MHz. For trap frequencies closer to the 1MHz range, this is much smaller than the range of $10^{-4}$ infidelity considered in this paper, hence the approximation is justified.

For the stretch mode the potential energy of the system is
\be
V(z)=\frac{ke^2}{2z}+m\omega^2\left(z^2+\frac{z^4}{\xi^2}\right),
\ee
where $\pm z$ are the axial positions of the two ions. The equilibrium separation is given by $d_0=2z_0$ where $z_0$ is given by $V'(z_0)=0$, which yields
\be
-\frac{L^2}{2z_0^2}+2 \frac{z_0}{L}+4\frac{z_0^3}{\xi^2 L}=0.
\ee
where $L=(ke^2/m\omega^2)^{1/3}$ is the length scale of the ion-separation in a harmonic trap \cite{james_quantum_1998}. %\mode{{\bf Missing citation here}}. 
This equation is solved numerically for obtaining the equilibrium position $z_0$. 
\begin{comment}
\be
-\frac{L \,l}{2z_0^2}+2 \frac{z_0}{l}+4\frac{z_0^3}{\xi^2 l}=0.
\ee
where $l=\sqrt{\hbar/m \omega}$ is the harmonic oscillator length and $L$ the range of the Coulomb interaction defined by $L=ke^2/\hbar\omega$.
\end{comment}
Let $u$ be the small displacement in the stretch mode, i.e., the positions of the ions are $z=\pm(z_0+u)$, we expand the potential $V(z)$ up to the fourth order in $u$ to obtain
\be
\frac{V(u)}{\hbar\omega}=\left(1+\frac{L^3 }{2z_0^3}+\frac{6 z_0^2}{\xi^2}\right)\left(\frac{u}{l}\right)^2-\left(\frac{L^3 l}{z_0^4}-\frac{4z_0 l}{\xi^2}\right)\left(\frac{u}{l}\right)^3+\left(\frac{L^3 l^2}{2 z_0^5}+\frac{l^2}{\xi^2}\right)\left(\frac{u}{l}\right)^4,
\ee
where $l=\sqrt{\hbar/m \omega}$ is the harmonic oscillator length.

\begin{comment}
\be
\frac{V(u)}{\hbar\omega}=\left(1+\frac{L\, l^2 }{2z_0^3}+\frac{6 z_0^2}{\xi^2}\right)\left(\frac{u}{l}\right)^2-\left(\frac{L\, l^3}{z_0^4}-\frac{4z_0 l}{\xi^2}\right)\left(\frac{u}{l}\right)^3+\left(\frac{L\, l^4}{2 z_0^5}+\frac{l^2}{\xi^2}\right)\left(\frac{u}{l}\right)^4.
\ee
\end{comment}
As the total energy is $m\dot u^2+V(\eta_z)$, the effective potential for the motion of the stretch mode is $V_{\text{eff}}(u)=V(u)/2.$ Using perturbation theory to estimate the anharmonic shift for this potential is not straightforward as the second order contribution of the cubic term can dominate the first order contribution of the quartic term. Therefore we compute the lowest three energies numerically with the Numerov method, and obtain the anharmonicity shift by $\chi=(E_2-E_1)-(E_1-E_0).$ 

Figure \ref{fig:stretch} shows the anharmonicity for two Yb$^{+}$ ions with mass $m=171$u in a potential with trapping frequencies, $\omega/2\pi$, varying from $0.1$ to $10$MHz and characteristic lengths, $\xi$, from $0.1$ to $1000\mu$m. There are two distinct contributions to the anharmonicity: the intrinsic anharmonicity of the Coulomb potential which increases with $\omega$, and the external anharmonicity from the trapping potential which increases with $1/\xi^2$. In the purely harmonic limit where $\xi$ is very large, anharmonicity $\chi \simeq 2\pi \times 100$Hz is achieved for $\omega \simeq 2\pi \times 5$MHz. The required trap frequency can be reduced with an addition of a small quartic component in the potential. Data from Fig.~\ref{fig:stretch} show that a potential with $\omega= 2\pi \times 100$kHz and characteristic length $\xi= 1\mu$m yields an anharmonicity of $\chi \simeq 2\pi \times 1$kHz.

To demonstrate that this range of parameter values is achievable with current trap design, the electrostatic potential for our trap design, with the layout shown in Fig.~\ref{fig:exp}a, is simulated. The control voltage on the electrodes is configured to create a quartic component in the potential. Fig.~\ref{fig:exp}b shows the potential obtained with BEM simulation. The trap frequency, $\omega\approx 2\pi \times 100$kHz, and the characteristic length, $\xi\approx 2\mu$m, corresponds to an anharmonicity $\chi\approx 2\pi \times 150$ Hz. For comparison, the anharmonicity for the COM mode in the same quartic potential is only $3\hbar/(4m\xi^2)\approx 2\pi \times 10$Hz.

\begin{figure}[t]
\includegraphics[width=0.35\columnwidth]{anh_stretch.pdf}
\caption{Dependence of the stretch mode's anharmonicity on the trap frequency and the characteristic length of the quartic potential.}\label{fig:stretch}
\end{figure}
\begin{figure}[t]
\includegraphics[width=0.6\columnwidth]{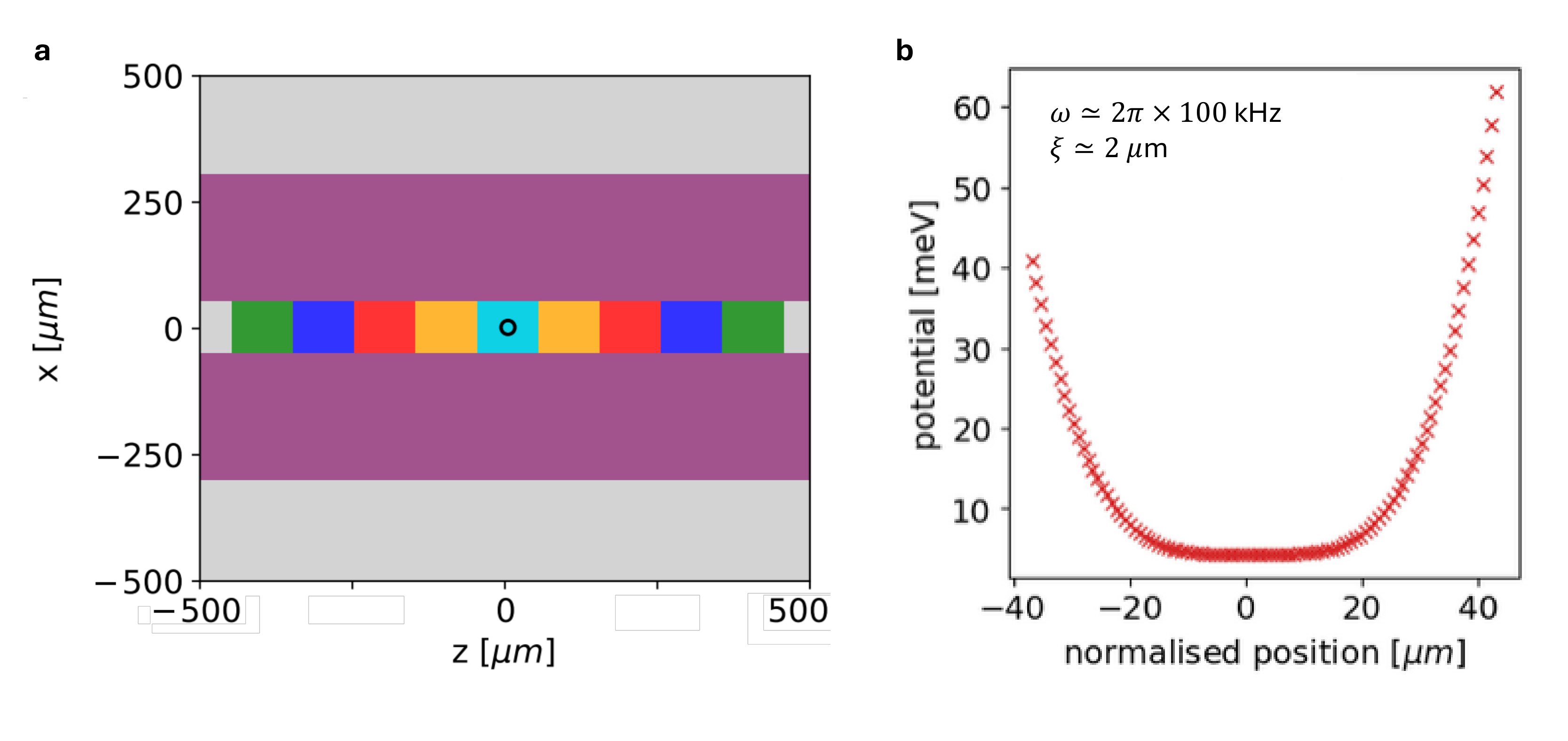}
\caption{a) Geometry of a surface ion trap with control electrodes placed directly underneath the central trapping axis, in between the RF electrodes (purple). Five different voltages are applied symmetrically to 9 control electrodes (cyan, orange, red, blue, green) to produce an axial potential with quadratic and quartic terms at the ion location (black circle). %\flo{Can we comment on what properties of the trap electrodes results in high anharmonicities?}
The quartic component can be enhanced by reducing the separations between the electrodes. b) Potential curve obtained with BEM simulation. The maximum voltage on the electrodes (outside the range shown in the figure)  is 70V.}\label{fig:exp}
\end{figure}

\subsection{2. Anharmonicity in a multi-ion chain}
In a chain with $n$ ions, the displacement of each ion in a normal mode is $z_j=z_j^{(0)}+u_j q$ where $z_j^{(0)}$ is the equilibrium position, $u_j$ the coefficient of the normal mode's eigenvector and $q$ the mode's spatial displacement \cite{james_quantum_1998}. The quartic term  in the total potential energy of the mode is
\be
 \alpha \frac{m\omega^2}{2} \frac{q^4}{\xi^2},
\ee
where 
\be
\alpha=\sum_j u_j^2,
\ee
determines the scaling with the system size.

For the COM mode $u_j=1/\sqrt{n}$ and hence the anharmonicity decreases as $1/n$. For the stretch mode,  
\be
u_j=\frac{z_j^{(0)}}{\sqrt{\sum_j \left(z_j^{(0)}\right)^2}},
\ee
which can be determined numerically. Using the numerical values for the equilibrium positions, $z_0^j$, from Ref.~\cite{james_quantum_1998}, we plotted $\alpha$ versus the system size, $n$, in Fig.~\ref{fig:alpha}, showing that for large system size $\alpha$, and hence the anharmonicity, scales as $1/n^\beta$ where $\beta\approx 0.9$. As with the previous approaches based on 
light-matter nonlinearity which scales as $1/n$, the robust gate is more useful in architectures where a few ions can be effectively trapped together, such as the quantum charge-coupled deivce (QCCD) architecture \cite{pino_demonstration_2021}.

\begin{figure}[t]
\includegraphics[width=0.4\columnwidth]{fig_alpha.pdf}
\caption{Dependence of the coefficient $\alpha$ with the number of ions in a chain (blue diamond). The fit (dashed line) shows a scaling $1/n^{\beta}$ where $\beta \approx 0.9$. \label{fig:alpha}}
\end{figure}

\subsection{3. Hamiltonian in the strong anharmonicity limit}

 The control Hamiltonian in the rotating frame of the free spin terms and the harmonic motional term, is then
 
\begin{equation}\label{eq:Hcapp}
\tilde{H}(t)\approx\sum_{n=0}^{\infty}\Delta_n\dyad{n} + \Omega_R\left[\tilde{f}^*(t)a+a^{\dagger}\tilde{f}(t)\right]S_y.
\end{equation}

The interaction Hamiltonian in the rotating frame of $\sum_{n}\Delta_n\dyad{n}$ is
\begin{align}\label{eq:Hint}
\tilde{H}_1(t)\approx \,&\Omega_R\sum_{n=0}^{\infty}\sqrt{n}\Big[\tilde{f}^*(t)e^{-i(\Delta_n-\Delta_{n-1})t}\dyad{n-1}{n} \nonumber \\ 
&+\dyad{n}{n-1}\tilde{f}(t)e^{i(\Delta_n-\Delta_{n-1})t}\Big]S_y.
\end{align}

In the strong anharmonicity limit where the anharmonicity is much larger than the Rabi frequency of the drive, we consider a polychromatic control of the form
\be\label{eq:polycontrol}
\tilde{f}(t)=\sum_{n=1}^{N-1} g_n(t)e^{-i(\Delta_{n}-\Delta_{n-1})t},
\ee
where $g_n(t)$ are slowly varying on the time scale $1/(\Delta_{n}-\Delta_{n-1})$. The control Hamiltonian reads, after neglecting the counter rotating terms,
\be\label{eq:Hrwa2}
\tilde{H}_2(t)\approx\Omega_R\sum_{n=1}^{N}\sqrt{n}[g_n^R(t)X_n-g_n^I(t)Y_n],
\ee
where $g_n^R(t)$ and $g_n^I(t)$ are the real and imaginary parts of $g_n(t)$. This rotating wave approximation is valid when $\chi\gg g_n(t)$, and only the $N$ lowest Fock states are involved in the dynamics.

\subsection{4. Required anharmonicity}

The required anharmonicity for achieving a sufficiently low average infidelity in the error range increases with increasing error magnitude. Figure~\ref{fig:reqanh} shows the minimum anharmonicity needed for achieving an average infidelity below $10^{-3}$ for the cases of 0 phonon, 1 phonon and 10 phonon excitations in the initial state. The anharmonicity rises with the error magnitude as expected but this rise is only moderate. 
\begin{figure}[h]
\includegraphics[width=0.4\columnwidth]{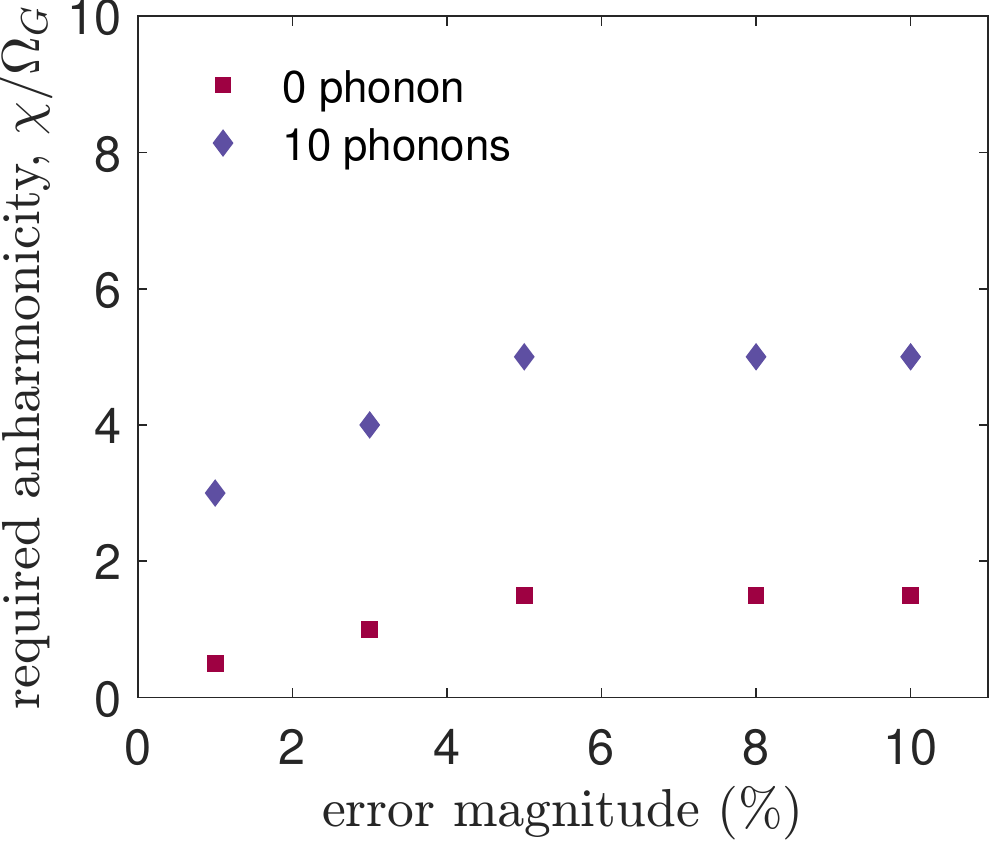}
\caption{Required anharmonicity for achieving an average infidelity of $10^{-3}$ w.r.t. error magnitude.}\label{fig:reqanh}
\end{figure}

\subsection{5. Numerical propagation and optimisation}

We used piece-wise controls and thus the control variables are the set of amplitudes for each time bin. The Hamiltonian Eq.~\eqref{eq:Hcapp} is written in matrix form using the basis $\ket{n}\otimes \ket{j_1}\otimes \ket{j_2}$ where $\ket{n}$ is the motional Fock state and $\ket{j_{1,2}}$ are the spin states. We include a sufficiently large number of Fock states in the basis to make sure there is no error caused by this truncation. This is verified by ensuring the population of the highest Fock state in the basis is negligibly small ($<10^{-10}$) at all time during the dynamics. The unitary evolution for each small time interval is obtained by matrix exponentiation with Padé approximation implemented in the expm function in Matlab. 

The gradient of the fidelity with respect to the control variable is computed using the analytical formula utilized in GRAPE \cite{khaneja2005optimal}. The fidelity is optimised using a gradient-based optimisation method implemented in the fmincon function in Matlab Optimization toolbox. 

We start with the strong anharmonicity limit where we find the optimal $g_n$, and use the solution in Eq.~\eqref{eq:polycontrol} as an initial guess to find the optimal $\tilde{f}(t)$ for lower anharmonicity, for which the numerical propagation must be done using the full Hamiltonian given in Eq.~\eqref{eq:Hcapp}.

\end{document}